# Improvement of Simplified Models of Variability of Stars: A review

Ivan L. Andronov

(1) Department "Mathematics, Physics and Astronomy", Odessa National Maritime University, Mechnikova, 34, 65029 Odessa, Ukraine, *tt_ari@ukr.net*

**Abstract:** Astronomical data are typically irregular in time, e.g. the space (HIPPARCOS/TYCHO, KEPLER, GAIA, WISE etc.) and ground-based CCD (NSVS, ASAS, CRTS, SuperWASP etc.) and photographic (Harvard, Sonneberg, Odessa etc.) photometrical surveys. This leads to cancellation of the conditions, which lead to the orthogonality of the basic functions, and thus the simplified methods give biased parameters of the approximations. In the common methods, there is a "matrix-phoebia", as it was later called by Prof. RNDr. Zdeněk Mikulášek, CSc. (MUNI).

We have elaborated a series of algorithms and programs for statistically correct analysis, and have applied them to 2000+ variable stars of different types. The data were obtained from an international collaboration in a frame of the "Inter-Longitude Astronomy" (ILA) campaign. Some highlights of our studies are presented, with an extended list of our original publications.

The main improvements were done: 1) for the periodogram analysis - the parameters are determined from a complete set of equations containing the (algebraic polynomial) trend superimposed on the (multi-) harmonic wave, so no "detrending", no "prewhitening" are used; 2) for the approximations - we use additional (multi-) harmonic waves, and also "special shapes" (patterns) for parts of the light curve, which correspond to relatively fast changes (minima of the eclipsing binaries, minima and maxima for the pulsating variables); 3) "auto correlation analysis" (ACF) - taking into account the bias due to a trend removal (previously - only a subtraction of the sample mean was taken into account); ACF for the irregularly spaced data; 4) for the signals with bad coherence, the "scalegram" analysis is proposed, which allows to estimate a characteristic cycle length and the amplitude, as well as to provide a realistic approximation; 5) the extension of the Morlet-type wavelet for more periodic signals and 6) "running" (parabola, sine) approximations for aperiodic and "nearly periodic" variations, respectively.

## Introduction

Time series analysis (recently more often called the "Data analysis") is applied not only in astronomy, but also in geoscience, economics and other sciences. Among the best textbooks are the ones by Anderson () The complexity of time series may vary from object to object and especially from one type of variable stars to another.

The majority of the methods of the time series had been elaborated for some standard cases of temporal behaviour. For real cases of multi component variability, some authors apply simplified (or even "oversimplified") methods, neglecting apparent correlations between different components of variability. This leads to shifts of values of the model parameters, and, consequently, may lead to wrong physical conclusions.This is generally true, even for linear model, with a summation of different contributions of variability, due too the orthogonality of the basic functions.In this case, the matrix of normal equations becomes non-diagonal, and the estimated values of the parameters depend on the number of coefficients. Obviously, this challenges their physical meaning.

We show some examples of wrong values of the coefficients and thus a mathematical model leading to wrong physical conclusions.

The basic equations started from Carl Gauss two centuries ago, and are published many textbooks and monographs (e.g. Anderson, Terebizh) and reviews (Andronov, 1994, 2003, Andronov & Marsakova, 2006, Mikulášek, 2007, 2015, Mikulášek et al., 2006, 2015). There may be some improvements, which are discussed below.

**Least Squares**

Let the approximation $x_C(t; C_\alpha)$ be the function with the corresponding parameters $C_\alpha, \alpha = 1..m$ are, which are to be determined.

In the vector notation, the test function is

$$\Phi(C_\alpha; \vec{t}; \vec{x}) = (\vec{x} - \vec{x}_C) \cdot (\vec{x} - \vec{x}_C), \tag{1}$$

where $\vec{x} = (x_k, k = 1..n)$ is the vector of observations obtained at times $\vec{t}$, and $\vec{x}_C = (x_{Ck}) = (x_C(t_k))$ is the vector of calculated values at times $t_k$. The best "quality of the approximation" corresponds to the minimal value of $\Phi$. Assuming an $n-$ dimensional distribution of the error of observations, this corresponds to a maximum likelihood. As the data are fixed, the test function $\Phi(C_\alpha; \vec{t}; \vec{x})$ is to be minimized in respect to the parameters $C_\alpha$ only. The generalized scalar product may be wrtten as

$$(\vec{a} \cdot \vec{b}) = \sum_{k,j=1}^{n} g(t_j, t_k, f; T_0) h_{jk} a_j b_k, \tag{2}$$

where $h_{jk}$ is an analog of the metric tensor, and $g$ is the function to make an approximations both local and dependent on the time scale and position of the central point (like in the wavelet and "running" approximations) (Andronov, 1997, 1998).

Mikulášek et al. (2003) proposed an additional weight function dependent on $|x_k - x_{Ck}|$, the aim of which is dump the outliers. In this case, the iteration is "robust" and should converge after several iterations.

Assuming that the errors of the measurements have a covariation matrix $\mu_{jk}$, it is usually recommended to use $h_{jk} = \sigma_0^2 \mu_{jk}^{-1}$, where $\sigma_0^2$ is any positive constant, which is called "the unit weight variance". The inverse relation is $\mu_{jk} = \sigma_0^2 h_{jk}^{-1}$.

The simplest case of variability is the "linear model"

$$x_C(t; C_\alpha), = \sum_{\alpha=1}^{m} C_\alpha f_\alpha(t), \tag{3}$$

where $f_\alpha(t)$ are basic functions, and the coefficents $C_\alpha$ are generally dependent on $f$ and $T_0$, if the additional weight function $g$ is not constant with time.

Practically, such complicated expressions are not used for the periodogram analysis. The matrix is usually oversimplified to $h_{jk} = \delta_{jk}$ (e.g. in the approximations in the electronic tables), or to $h_{jk} = w_k \delta_{jk}$, if assuming that $w_k = \sigma_0^2/\sigma_k^2$, $\mu_{jk} = \sigma_k^2 \delta_{jk}$. Such approximations are used typically for "global" approximations, when all the data are used.

The cases of overlapping "running" approximations (improving the "running mean" = "moving average") are discussed below.

**Periodogram analysis**

"Point – Point" Methods

The structural scheme of the methods of the periodogram analysis was presented by Andronov (1996). The main division is into the groups "point-point" (non-parametric) and "point-curve" (parametric). The first group has started from Lafler and Kinman (1965) and was modified by other authors. All these methods are based on minimizing weighted mean distance (or its square) between the points with times $t_k, k = 1..n$, values of the signal $x_k$, which are sorted according to phases $\phi_k$ computed for a given trial period $P$ (or frequency $f = 1/P$). The test function may be generally expressed as a function of trial frequency and fixed (for a given sample) data:

$$\Theta(f; t_k; x_k) = A \sum_{k=1}^{n} \rho(|x_k - x_{k-1}|; |\phi_k - \phi_{k-1}|), \tag{4}$$

where the initial data $(x_k, \phi_k, k = 1..n)$ are extended by $x_0 = x_n, \phi_0 = \phi_n - 1$. The scaling coefficient $A > 0$. The summand $\rho(\delta_\phi; \delta_x)$ may be equal e.g. to $|\delta_x|^\gamma$ with a power index $\gamma = 1$ (appendix by T.J.Deeming to Bopp et al., 1970), $\gamma = 2$ (initial proposal by Lafler and Kinman, 1965), or any other positive value. Few variations of functions with a real dependence of $\rho$ on $\delta_\phi$ were proposed e.g. by Renson (1978) and Dworetsky (1983).

Detailed review and comparison of these methods was presented by Andronov and Chinarova (1997).

More general expression was presented in Eq. (3) by Pelt (1975) and may be reasonably rewritten using the "structure function".

$$\Theta(f; t_k; x_k) = \frac{\sum_{j=1}^{n-1} \sum_{k=j+1}^{n} g(f; |t_k - t_j|) \rho(|x_k - x_j|)}{\sum_{j=1}^{n-1} \sum_{k=j+1}^{n} g(f; |t_k - t_j|)}, \quad (5)$$

where $\rho(\delta_x)$ is the "distance" between the points (e.g. again $|\delta_x|^\gamma$), and the structure function may be split into two parts:

$$g(f; |t_k - t_j|) = W(f; |t_k - t_j|) G(|\phi_k - \phi_j|), \quad (6)$$

where, in the simplest case, function $W(f; \delta_t) = 1$, if $|\delta_t| \leq \delta_{t,\text{lim}}$ (else $W(f; \delta_t) = 0$), and, similarly, $G(\delta_\phi) = 1$, if $|\delta_\phi| \leq \delta_{\phi,\text{lim}}$ or $1 - |\delta_\phi| \leq \delta_{\phi,\text{lim}}$ ($G(\delta_\phi) = 0$). In other words, the structure function is a sum of $\rho(|x_k - x_j|)$ only for the data points close either in phase, or in time. Another improvement may be of a "wavelet – style", if $W(f; \delta_t) = 1$, if $|f\delta_t| \leq \delta_{E,\text{lim}}$. One may propose smooth functions, e.g. a Gaussian $W(f; \delta_t) = \exp(-c((f\delta_t)^2)$ or a limited-width function proposed by Andronov (1997). For the second function, Pelt (1975) suggested a cosine-type shape $G(\delta_\phi) = (1 + \cos(2\pi\delta_\phi))/2$, which is a good choice for nearly sinusoidal oscillations, but is much worse for periodic variables with more complicated curves (e.g. eclipsing binaries). Because of long computational time needed, these methods are not widely used, except the initial method of Lafler and Kinman (1965). It was realized in many programs (e.g. Breus, 2003, 2007; Vanmunster, 2018).

"Point – Curve" Methods

The "point-curve" methods are typically based on the least squares method by comparing the data with the approximation ("computed curve") $x_C(t; C_\alpha)$, which is dependent on the frequence $f = \frac{1}{P}$, where $P$ is a trial period, and (generally) on the initial epoch $T_0$. Obviously, the test function $\Phi(C_\alpha; \vec{t}; \vec{x})$ depends on these two parameters as well. The simplest type of variability is a cosine wave. For it, the optimal mathematical model is:

$$x_C(t; C_\alpha) = C_1 + C_2 \cos \omega t + C_3 \sin \omega t = C_1 - R \cos(\omega(t - T_0)), \quad (7)$$

where $\omega = 2\pi f = 2\pi/P$, and $T_0$ is the "initial epoch" corresponding to a *minimum* signal value (which, if being a magnitude (in astronomy), as a *maximum* of brightness). The relations between the coefficients are well known: $C_2 = -R \cos(\omega T_0)$, $C_3 = -R \sin(\omega T_0)$, $R^2 = C_2^2 + C_3^2$, $T_0 = \arctan2(-C_3, -C_2) + P \cdot K$, where $K$ is any integer number. The function arctan2(y,x) calculates arctan(y/x), and returns an angle in the correct quadrant (is present in Python, Delphi and other computer languages). For further period corrections, Andronov (1994) recommended to use such a value of $K$, which makes $T_0$ as closest as possible to a (weighted) sample mean of times of observations.

For mono-periodic multi-harmonic signals (e.g. "regular" pulsating variables and eclipsing binaries), this may be easily expanded to trigonometric polynomials (abbreviated Fourier series) of order $s$:

$$x_C(t; C_\alpha) = C_1 + \sum_{j=1}^{s} (C_{2j} \cos j\omega t + C_{2j+1} \sin j\omega t) = C_1 - \sum_{j=1}^{s} R_j \cos(j\omega(t - T_{0j})) \quad (8)$$

with similar relations: $C_{2j} = -R_j \cos(j\omega T_{0j})$, $C_{2j+1} = -R_j \sin(\omega T_{0j})$, $R_j^2 = C_{2j}^2 + C_{2j+1}^2$, $T_{0j} = \arctan2(-C_{2j+1}, -C_{2j}) + \frac{P \cdot K}{j}$.

The test function is recommended to be computed at a grid of test frequencies $f_i = f_0 + i \cdot \Delta f$, where $\Delta f = \Delta \phi/(t_n - t_1)$, and the phase shift is recommended to be $0 < \Delta \phi \leq 0.1$ (cf. Andronov, 1994b). After determination of the largest peak at the periodogram, the period may be corrected by differential corrections. The realizations of the method were made initially in the Fortran programming language, but then implemented in the MCV ("Multi-Column Viewer") by Andronov and Baklanov (2004).

Examples of periodograms for different degrees $s$ and discussion on the statistically optimal degree of the trigonometrical polynomial are presented by Andronov et al. (2016).

Based on the statistically optimal fits of a group of Mira-type stars, Kudashkina and Andronov (2017) compiled an atlas of behavior of the light curves at the $(x_C; dx_C/d\phi)$ phase diagrams.

**Multi-Periodic Multi-Harmonic Oscillations with Trend:**
**Detrending / Prewhitening vs Complete Models**

Similarly to multi-harmonic approximations, the fit may be expanded to few ($L$) independent periods $P_l, l = 1..L$, with corresponding degrees of the trigonometrical polynomial $s_l$, and the possible trend may be represented by an algebraical polynomial of degree $s_0$:

$$x_C(t; C_\alpha) = \sum_{j=1}^{s_0+1} C_j \cdot (t - T_{mean})^{j-1} - \sum_{l=1}^{L} \sum_{j=1}^{s_l} R_{jl} \cos(j\omega_l \cdot (t - T_{0jl})) \qquad (9)$$

In the software MCV (Andronov and Baklanov, 2004), the maximal number of independent periods is $L = 3$. For the periodogram analysis, $L = 1$, but one may use a trend and a multi-harmonic approximation. The test function is

$$S = 1 - \Phi_{s_0+1+2s_1}/\Phi_{s_0+1}, \qquad (10)$$

where $\Phi_m$ is the test function for the approximation with $m$ parameters. For simpler approximations (e.g. in the electronic tables etc.), the value of $S$ is expressed as $r^2$. This value shows the ratio of the variance of the "signal" to the "signal+noise", where the "signal" is in deviation of the "trend+periodic wave" approximation from the "pure trend".

Detrending is a special type of prewhitening, when the characteristic time scale is much longer than the observations, so an algebraic polynomial is used instead of of multi-harmonic approximation.

It is very important to note, that, due to the non-ortogonality of the basic functions, the coefficients $C_j$ of the "trend" are *generally different* from that of the "trend+periodic wave". In the simplified methods, the "detrending" is applied, and then the periodogram analysis is applied as for the signal without any trend. This biases the periodogram and may produce peaks at false periods (frequencies), what produces *fake results*. Similarly, "prewhitening" is a subtraction of a periodic wave, and further application of the periodogram analysis. For a "good" distribution of times of the signal (approximately filling a complete phase curve), this may not be very significant. But for real observations of superhumps or pulsations of the δ Sct-type stars, the periods of which are comparable with the duration of the observations, the errors may reach dozens of per cent.

In the light curves of the intermediate polars, there are two periodic components (orbital and spin, e.g. Patterson, 1994), thus it is effective to apply (e.g. Andronov and Breus, 2013). Andronov et al. (2008a) used trigonometrical polynomials to study night-to-night variations of the asynchronous polar BY Cam.

The total number of parameters should not exceed a default value of $m = 21$. Of course, all the parameters are computed with corresponding error estimates, as defined in the LSQ approximation.

**Special Shapes for the Narrow Features of the Signal**

Polynomial Splines

Trigonometrical polynomial fits are excellent approximating functions in a case of smooth curves. However, in a case of abrupt variations (eg. narrow eclipses in binary stars, or relatively short ascending branches of RR Lyr - type stars or HADS (high amplitude δ Scuti stars)), the number of the parameters becomes very large, causing apparent waves at the light curve (the Gibbs phenomenon). In this case, it is recommended to use a smaller number of basic functions.

In the case of one hump phase curves of pulsating variables, we have applied a two-interval model consisting of a parabola at the longer part and a cubic function in the smaller interval typically corresponding to a rapidly ascending branch. The function itself, and its derivative are continuous everywhere, including the borders between the intervals. There are 3 parameters obtained using linear LSQ, and 2 nonlinear parameters, including two positions of borders.

These cyclic spline approximations of variable order 2 and 3 were used for a period search and parameter determination of the pulsating variables from the HIPPARCOS-TYCHO photometrical survey (Andronov, Cuypers & Piquard, 2000).

Another approximation, based on splines of different order, is a "constant+parabola" one. There are also 3+2 parameters. Such an approximation is an effective one not only for the EA-type stars, but also for EB and EW . The position of the minimum and the eclipse width are the free parameters (Andronov, Cuypers & Piquard, 2000). This approximation has much better quality than e.g. the approximation in 4 overlapping intervals, proposed by Papageorgiou et al. (2014), which is good for the maxima (out-of-eclipse) phases, but has systematic underestimate of the eclipse depth.

Cubic splines with periodic bound conditions were used by Andronov (1987) for the periodogram analysis and studies of the period variations without using the (O-C) diagrams for the times of extrema, using a trial "time correction" to compute phases taking into account not only the initial epoch and period, but a term proportional to the period derivative. As the splines are dependent not only on the number of basic points, but also on the shift, there were two versions – the "shift-averaged" spline and the "shift-optimized" spline.

Local Special Shapes (Patterns)

Further improvements are due to using other functions rather than algebraic or trigonometric polynomials. Andronov (2010) proposed to use a combination of the trigonometrical polynomial of only second order with "eclipse parts" with a local "special shape" $H(z) = (1 - |z|^\alpha)_+^{1.5}$, where the index "+" denotes that negative values of the function are set to zero. The function may be scaled and shifted to fit the minimum, and the parameter $\alpha$ describes the shape at the center. The minimal (physically reliable) value $\alpha = 1$ corresponds to a total eclipse of stars with equal radii, and $\alpha \to \infty$ corresponds to an extremely "flat" minimum. The fixed power index 1.5 corresponds to asymptotic behavior of the light curve close to the phases of the outer contacts at the eclipse (Andronov and Tkachenko, 2013).

This approximation was referred as to the NAV ("New Algol Variable") one, and was initially appled to four newly discovered variable stars by Kim et al. (2010) and then to other "newbees". Andronov (2012a) argued that this local approximation seems the best among others with the same number of parameters because, with the same quality of the fit, it allows to determine, in an addition to other parameters, an important characteristic for the "General Catalog of Variable Stars" (Samus' et al., 2017) – the eclipse full half-width, Unfortunately, this important parameter (making a relation between the summary dimensionless size of the components and the inclination) is recently typically ignored, as the popular programs do not determine it.

This parameter is, in principle, may not be determined using functions with formally infinite width, e.g. a Gaussian or its "hyperbolic cosine" improvement by Brat et al. (2012), implemented as a standard on-line tool at the B.R.N.O. (O-C gate) web page. For further details, see Mikulášek (2015) and Mikulášek et al. (2015),

Andronov (2012b) compared the "NAV" function to others, including the function $H_2(z) = (1 - z^2)_+^\beta$, which has a classical parabolic shape at the center $|z| \ll 1$: $H_2(z) \approx (1 - \beta z^2)$, contrary to the "NAV" shape with another $H(z) \approx (1 - 1.5 \cdot |z|^\alpha)$. At the eclipse borders, $|z| \approx 1$, $H(z) \approx \alpha^{1.5}(1 - |z|)^{1.5}$, and $H_2(z) \approx 2^\beta(1 - |z|)^\beta$ – both functions have power shape, but different. The function $H_2(z)$ was applied to 26121 LMC and 6138 SMC eclipsing binaries from the OGLE III LMC database (Graczyk et al., 2011) photometric survey by Juryšek et al. (2018).

The "New Algol Variable" (NAV) algorithm may be effectively applied not only to Algol-type stars, as was initially suggested, but also to β Lyr and even W UMa stars. The curves for the prototype stars of these subclasses are presented by Tkachenko et al. (2016).

Contrary to the approximations of complete light curves, the phenomenological modeling of the partial light curves covering only the vicinities of extrema (to determine only the ToM=Time of Minimum/Maximum) is a more frequent task. The previous review on the statistically optimal determination of the characteristics of extrema (not only ToM, but the approximating function as well) was presented by Andronov (2005).

The polynomial of a statistically optimal order was used for compiling the catalogue of the characteristic of individual extrema of 173 semi-regular variables by Chinarova and Andronov (2000). This method was later implemented by Breus (2003).

For the intervals, completely covering the eclipses, Andronov et al. (2017) proposed few improvements of the methods proposed be Andronov (2012a) and Mikulášek (2015). For the intervals, covering the symmetric eclipses, or maxima of the pulsating variables, Andrych et al. (2015, 2017) proposed a series of which split the interval into two or three parts. These more recent methods were implemented in the software MAVKA, and are presented in the same volume by Andrych and Andronov (2018). The first version (Andronov et al., 2015) implemented the approximations using (general and symmrtric polynomials as well as the asymptotic parabola (Marsakova and Andronov, 1996).

Andrych et al. (2017) introduced the "wall-supported" functions which are more effevtive for total eclipses and transits. Finally, the 3-interval parabolic spline was implemented.

The program MAVKA allows to apply up to 20 functions for a given interval and may automatically choose the best approximation, which cortesponds to the best accuracy of the ToM.

Previous versions of the program were applied to observations of some eclipsing binary stars (Savastru et al, 2017; Tvardovskyi et al., 2017, 2018) and symbiotic variables (Marsakova et al., 2015).

An alternate approach is to determine moments of crossing of the approximation and some constant level (cf. Andronov et al., 2008b, Andronov and Andrych, 2014).

### "Running" Approximations

Thus the one may use $T_0$ as an argument of the "super-function", which is constructed from the middle points of the original functions computed by taking into account the Eq. (6):

$$\widetilde{x_C}(T_0), = \sum_{\alpha=1}^{m} C_\alpha(f; T_0) f_\alpha(T_0). \qquad (11)$$

A complete theory of the statistical properties of such functions and their derivatives was presented by Andronov (1997) with an application to "running parabola" (RP) with an additional weight function

$$g(t_j, t_k, f; T_0) = (1 - ((t_j - T_0)f)^2)_+ \cdot \delta_{jk}. \qquad (12)$$

Here the bottom index "+" denotes that the value is computed in this way, only if the result is positive (else $g = 0$). The statistically optimal value of $f = 1/\Delta t$ (in the original paper) may be determined using the scalegram analysis

The application to the sine functions leads to two branches: the running sines and the wavelet analysis.

### Running Sines (RS)

For the "Running Sines" (RS), the function is defined by Eq. (1), and is an effective tool for (nearly) periodic functions with a (relatively slow) trend, e.g. semi-regular variables, intermediate polars. The review on the "stress test" of the method for the signals with variable periods was presented by Andronov and Chinarova (2013).

Running sines are an intermediate between the wavelet and a running parabola, discussed below. Contrary to the wavelet, the period is fixed, as well as as the filter half width. The weight functions is constant inside the interval of smoothing, i.e. the filter (weight) function is rectangular.

The recommended filter half-width is $\Delta t = P/2$, but the pulsational periods of many Mira-type stars are close to a year, thus seasonal gaps in the observations cause large errors if the smoothing function, thus, for these signals, the recommendation is $\Delta t = P$. Generally, it is as a free parameter. Larger values of $\Delta t$ may be recommended for relatively stable light curves, in which the shape of the light curve changes at a time scale exceeding $\Delta t$, so will hide possible real variations of the characteristics of the light curve.

Theoretically, for pure multiharmonic signal and an interval, uniformly filled with observations, the smoothing function is a purely sinusoidal function. For the data with gaps, the shape becomes a-sinusoidal.

**Wavelet Analysis**

Wavelet analysis is a popular tool for analyzing nearly periodic data with time scale of variability of characteristics of the individual cycles, which exceeds the period by a factor of a few times. The over-simplified method is based on a simple replacement of the integrals from minus to plus infinity by discrete sums over the moments of observations, similarly to the method of Deeming (1975). Foster (1995) proposed the WWZ ("weighted wavelet Z – transform"), which is based on a determination of the parametrrs of the fit using the least squares method with weights. Andronov (1998) proposed improvements of the method, introducing the wavelet periodogram, and using for the period determination the test function $S(f)$ similar in sense to that one, which was introduced by Andronov (1994a).

Depending on the time distribution of the observations, the improvement of the "signal-to-noise" ratio may reach "gain" from dozens of per cent to dozens of times (for a very disjointed distribution in time).

The recommendation for the wavelet analysis (based on the improved Morlet-type wavelet) is to use the logarithmically constant scale step for the trial periods, contrary to usually used periods. Also as an analog of the Nyquist interval between the data, for the Morlet type wavelet, the optimal time step for smoothing is $P/3$.

Detailed reviews were presented by Andronov (1998, 1999).

**Running Parabolae**

This method was elaborated by Andronov (1997) for irregular signal and the signals of low coherence, e.g. quasi-periodic oscillations (QPO) or regular outbursts with variable occurrence rate. The free parameter to be determined for the statistically optimal smoothing is the filter half-width $\Delta t$. Similarly to the periodogram analysis, the scalegram analysis is used. The corresponding test functions may be different. Among them - the unbiased estimate of r.m.s. deviation of the observations from the fit; the r.m.s. accuracy of the approximation at the arguments of the signal; the "signal-to-noise" ratio. The weight function (in the notation of the Eq. (12)) is:

$$g(t_j, t_k, f; T_0) = (1 - ((t_j - T_0)/\Delta t)^2)_+^2 \cdot \delta_{jk} \qquad (13)$$

Andronov (2003) introduced an additional test function $(\Delta t)$, , which may be interpreted as a "scalegram periodogram". It may be used for determination of "period" (cycle length" of quasi-periodic signals (QPO) and effective amplitude. Obviously, for pure sinusoidal signals with "good" (dense, uniform) distribution of the times of observations, all these methods (global fit, running sines, wavelet analysis and the scalegram analysis) should result in the same (within accuracy estimates) parameters. However, for real data, the results may be significantly different. Also, the scalegram analysis an effective tool for the flickering (red noise, fractal) variability, e.g. in cataclysmic and symbiotic variables.

**Autocorrelation Functions (ACF) of Detrended Signals**

Autocorrelation analysis is widely used for the analysis of regularly spaced time series i.e. when the times of the observations are related as $t_k - t_j = (k - j) \cdot \delta$ (e.g. Anderson, 1971).

This classical relation may be improved for the ACF analysis of the residuals of the observations from the smoothing function, what is an often case e.g. for cataclysmic variables with orbital variability and quasi-periodic oscillations (QPO) and/or flickering, as well as for binary systems with pulsations of one of the components

The removal of the mean was studied by Sutherland et al. (1978), whereas the complete theory of statistical properties of the ACF was presented by Andronov (1994a). This may help to prevent false interpretation of flickering as the quasi-periodic oscillations.

**Acknowledgement**

We thank Lewis M. Cook and Prof. RNDr. Zdeněk Mikulášek for discussions. This work is related to the international projects "Inter-Longitude Astronomy" (ILA, Andronov et al., 2003, 2014, 2017) and "Ukrainian Virtual Observatory", "Astroinformatics" (Vavilova et al., 2017).